\documentclass[11pt,preprint,showpacs]{revtex4}
\usepackage{graphicx}
\usepackage{dcolumn}
\usepackage{bm}
\begin{document}

\title{Low-energy neutron-deuteron reactions with N$^3$LO chiral forces}

\setcounter{page}{1}

\author{J.~Golak}
\affiliation{M. Smoluchowski Institute of Physics, Jagiellonian
University,  PL-30059 Krak\'ow, Poland}

\author{R.~Skibi\'nski}
\affiliation{M. Smoluchowski Institute of Physics, Jagiellonian
University,  PL-30059 Krak\'ow, Poland}

\author{K. Topolnicki}
\affiliation{M. Smoluchowski Institute of Physics, Jagiellonian
University,  PL-30059 Krak\'ow, Poland}

\author{H.~Wita{\l}a}
\affiliation{M. Smoluchowski Institute of Physics, Jagiellonian
University,  PL-30059 Krak\'ow, Poland}

\author{E. Epelbaum}
\affiliation{Institut f\"ur Theoretische Physik II, Ruhr-Universit\"at
  Bochum, D-44780 Bochum, Germany}

\author{H. Krebs}
\affiliation{Institut f\"ur Theoretische Physik II, Ruhr-Universit\"at
  Bochum, D-44780 Bochum, Germany}

\author{H.\ Kamada}
\affiliation{Department of Physics, Faculty of Engineering,
Kyushu Institute of Technology, Kitakyushu 804-8550, Japan}

\author{Ulf-G. Mei{\ss}ner}
\affiliation{Helmholtz-Institut f\"ur Strahlen- und
             Kernphysik and Bethe Center for Theoretical Physics, \\
             Universit\"at Bonn,  D--53115 Bonn, Germany\\}%
\affiliation{Institute~for~Advanced~Simulation, Institut~f\"{u}r~Kernphysik,
J\"{u}lich~Center~for~Hadron~Physics, and JARA~-~High~Performance~Computing
Forschungszentrum~J\"{u}lich,
D-52425~J\"{u}lich, Germany}

\author{V. Bernard}
\affiliation{Institut de Physique Nucl\'eaire, CNRS/Univ. Paris-Sud
  11, (UMR 8608), F-91406 Orsay Cedex, France}

\author{P. Maris}
\affiliation{Department of Physics and Astronomy, Iowa State
  University, Ames, Iowa 50011, USA}

\author{J. Vary}
\affiliation{Department of Physics and Astronomy, Iowa State
  University, Ames, Iowa 50011, USA}

\author{S. Binder}
\affiliation{Institut f\"ur Kernphysik, Technische Universit\"at 
Darmstadt, 64289 Darmstadt, Germany}

\author{A. Calci}
\affiliation{Institut f\"ur Kernphysik, Technische Universit\"at 
Darmstadt, 64289 Darmstadt, Germany}

\author{K. Hebeler}
\affiliation{Institut f\"ur Kernphysik, Technische Universit\"at 
Darmstadt, 64289 Darmstadt, Germany\\}
\affiliation{Extreme Matter Institute EMMI, GSI Helmholtzzentrum f{\"u}r
Schwerionenforschung GmbH, 64291 Darmstadt, Germany}

\author{J. Langhammer}
\affiliation{Institut f\"ur Kernphysik, Technische Universit\"at 
Darmstadt, 64289 Darmstadt, Germany}

\author{R. Roth}
\affiliation{Institut f\"ur Kernphysik, Technische Universit\"at 
Darmstadt, 64289 Darmstadt, Germany}

\author{A. Nogga}
\affiliation{Institut f\"ur Kernphysik, Institute for Advanced Simulation 
and J\"ulich Center for Hadron Physics, Forschungszentrum J\"ulich, 
D-52425 J\"ulich, Germany}

\author{S. Liebig}
\affiliation{Institut f\"ur Kernphysik and J\"ulich Center for Hadron
  Physics, Forschungszentrum J\"ulich, D-52425 J\"ulich, Germany}

\author{D. Minossi}
\affiliation{Institut f\"ur Kernphysik and J\"ulich Center for Hadron
  Physics, Forschungszentrum J\"ulich, D-52425 J\"ulich, Germany}

\date{\today}

\begin{abstract}

  We solve three-nucleon Faddeev equations with nucleon-nucleon and
  three-nucleon forces derived consistently in the framework 
of chiral perturbation theory 
 at next-to-next-to-next-to-leading order in the chiral expansion. 
 In this first investigation we include only matrix elements
of the three-nucleon force for partial waves
with the total two-nucleon (three-nucleon) angular momenta 
up to $3$ ($5/2$).
 Low-energy neutron-deuteron elastic scattering and deuteron breakup 
reaction are studied. 
  Emphasis is put on $A_y$ puzzle in elastic scattering and cross
  sections in  symmetric-space-star and neutron-neutron quasi-free-scattering 
breakup configurations,  for which large
  discrepancies between data and theory have been reported. 
\end{abstract}


\maketitle

\section{Introduction}
\label{intro}

A special place among
few-body systems is reserved for the three-nucleon (3N) system, for which
 a mathematically sound theoretical formulation in the form of Faddeev
equations exists, both for bound and scattering states. 
Over the past few decades  algorithms have been developed 
 to solve numerically  three-nucleon Faddeev equations
for any dynamical input which, in addition to nucleon-nucleon (NN)
interactions,  also involves 
three-nucleon forces (3NFs) \cite{wit88,glo96,hub97}. 
Using these algorithms and standard, (semi)phenomenological
nucleon-nucleon interactions alone or supplemented by three-nucleon
force model, numerous investigations of 3N bound states and reactions in 
 the 3N continuum have been carried out.   
High precision nucleon-nucleon potentials such as the AV18~\cite{AV18},
CD~Bonn~\cite{CDBOnucleon-nucleon}, Nijm I and II~\cite{NIJMI} NN forces,
which provide a very good description of the nucleon-nucleon data  up to
about 350 MeV, have been used.  
 They have also been combined with model 3N forces such as the $2\pi$-exchange
 Tucson-Melbourne (TM99) 3NF \cite{TM99} or the Urbana IX model \cite{uIX}.

When realistic NN  forces are used to predict binding energies
of three-nucleon systems they typically underestimate the experimental bindings
of $^3$H and $^3$He by about 0.5-1
MeV~\cite{Friar1993,Nogga1997}. This missing binding energy can be
corrected for by introducing a three-nucleon force into the nuclear
Hamiltonian~\cite{Nogga1997}. 
Also the study of elastic nucleon-deuteron (Nd) scattering and nucleon
induced deuteron breakup revealed a number of cases where the
nonrelativistic description using only pairwise forces is insufficient
to explain the data.  The best studied case at low energies is the
vector analyzing power in elastic Nd scattering for
which a large discrepancy exists in the region of its maximum around
c.m. angles $\theta_{c.m.} \sim 125^o$ and for incoming nucleon
energies below $\sim 20$~MeV \cite{glo96,wit01,kie_2001}.  
 For the elastic scattering angular
distribution at such energies, negligible effects of 3NF's have been
found and theory based on realistic NN forces agrees well 
with the data \cite{glo96,wit01}.

That picture changes with increasing energy of the 
Nd system.  
Generally, the studied discrepancies between experiment and 
theory using only NN  potentials  become
larger and adding a
three-nucleon force to the pairwise interactions leads in some cases
to a better description of the data.  The elastic Nd 
angular distribution in the region of its minimum and at backward
angles is the best known example~\cite{wit98,sek02}.  The clear
discrepancy in these angular regions at energies up to
$E_{\rm lab, \, N}\sim 100$~MeV between a theory using only
 NN potentials and the cross section data can be removed
by adding  standard models of  three-nucleon forces to the nuclear
Hamiltonian. Such 3NFs are adjusted for a given  
 NN potential to reproduce the experimentally observed binding energy of
$^3$H and $^3$He~\cite{wit98,wit01,sek02}.  At energies higher than
$\sim 100$~MeV current 3NFs only partially improve
the description of cross section data and the remaining discrepancies,
which increase with energy, indicate the possibility of relativistic
effects.  The need for a relativistic description of three-nucleon
scattering was  realized when precise measurements of the total
cross section for neutron-deuteron (nd) scattering~\cite{abf98} were
analyzed within the framework of nonrelativistic Faddeev
calculations~\cite{wit99}.  NN forces alone were
insufficient to describe the data above $\sim 100$~MeV.  The
effects due to the relativistic kinematics considered in \cite{wit99} 
at higher energies were comparable in magnitude 
 to the effects due to 3NFs. 
 These results provided further motivation to study 
relativistic effects in the three nucleon continuum in a systematic way.

Subsequent studies of relativistic effects in the three-nucleon continuum
\cite{witrel1,witrel2,rel3nf,erratarel3nf}  revealed, 
that when the non-relativistic form of the kinetic energy is replaced
by the relativistic form and a proper treatment of the relativistic dynamics is
 introduced, the elastic scattering cross section is only slightly
 increased at backward angles and higher energies 
 while spin observables are practically unchanged. 
 These results led to the conclusion that 
 discrepancies between data and theory at higher energies 
 must reflect the action of 3NF's
 which have to be included in the nuclear Hamiltonian.

The main drawback of all those studies was 
 inconsistency between
applied NN interactions and 3NFs. 
 With the advent of effective field theoretical methods in the form of
 chiral perturbation theory, it became possible to construct consistent
 two- and many-nucleon forces. 
 In this way  an exciting possibility  
to study few-nucleon
systems and their reactions with consistent two- and 
 many-nucleon interactions has emerged. 

In \cite{epel2002}, the above mentioned inconsistency 
was removed and low-energy 3N continuum were investigated
with chiral next-to-next-to-leading order (N$^2$LO) 
 NN and 3N forces. The NN interaction in that
order, however, does not describe the NN experimental  phase-shifts in
 a sufficiently wide energy range to allow application of those forces 
at higher energies.

In \cite{epel_nn_n3lo,epel_mod,epel_ham_meis} and
  \cite{mach_nn_n3lo,mach_phydrep}, 
precise two-nucleon potentials have been
developed at next-to-next-to-next-to-leading order (N$^3$LO) of the chiral
expansion. They reproduce experimental NN phase-shifts 
\cite{nijm_phase1,nijm_phase2} in a wide
energy range and practically with the same high precision as realistic
(semi)phenomenological NN potentials. The necessary work to derive the
 consistent chiral 3NF's at N$^3$LO has been 
accomplished in \cite{3nf_n3lo_long,3nf_n3lo_short} 
and \cite{rob_ishi}. At that
order, six different topologies contribute to the 3NF. Three of them
 are of a  long- and intermediate-range \cite{3nf_n3lo_long} 
 and are given by two-pion (2$\pi$)
exchange, two-pion-one-pion (2$\pi$-1$\pi$) exchange 
and  the so-called ring diagrams. They are
supplemented by the shorter-range 
 $1\pi$-contact and three-nucleon-contact components, which appear
 first  at N$^2$LO,  by the two-pion-exchange-contact (2$\pi$-contact)
 term as well as by the leading relativistic corrections to the
three-nucleon force \cite{3nf_n3lo_short}. 

The results of 
Refs.~\cite{mach_nn_n3lo,epel_nn_n3lo,epel_mod,3nf_n3lo_long,3nf_n3lo_short} 
  enable one to perform, for the first time, consistent
calculations of three-nucleon reactions at N$^3$LO order of chiral  
expansion.   The 3NF at this
order does not involve any new unknown low-energy constants (LECs)
and depends only on two free parameters, $c_D$ and $c_E$, which parametrize the
 strengths of the leading $1\pi$-contact  and the
 three-nucleon-contact terms. 
  Their values need to be fixed (at given order) 
 from a fit to a 
few-nucleon data. Among the few possible observables that have been
used in this connection one can mention the triton binding energy, the
 nd doublet scattering length $^2a_{nd}$ \cite{epel2002}, the
$^4$He binding energy \cite{kie_2010,navra_2007} along with the point
 proton rms radius \cite{navr2007b}, the properties 
 of light nuclei, or the triton $\beta$ decay
 rate~\cite{marcucci_beta}. 
 Notice that the
 first three observables are known to be strongly correlated and
 therefore might not be the best choice for the determination of
 $c_D$ and $c_E$. 

Application of N$^3$LO 3NF in few-body calculations is challenging due
to its very rich and complicated operator structure. The large number of
terms in the 3NF at N$^3$LO \cite{3nf_n3lo_long,3nf_n3lo_short}  
 requires an efficient method of
performing partial-wave decomposition. Recently such a method, which
 runs under the name of automatized partial-wave decomposition (aPWD), was
proposed  in \cite{apwd,apwd_a,apwd_a32a}. In that approach, 
 the matrix elements in the 3N momentum-space
partial wave basis  for different terms
contributing to N$^3$LO 3NF are obtained in two consecutive steps. 
First, the spin-momentum and isospin parts of three-nucleon
interactions are computed using symbolic algebra systems. 
The resulting momentum-dependent functions are then
integrated numerically in five dimensions over angular variables. The
major advantage of this method is its generality since it can be applied
to any momentum-spin-isospin operator. Application of that method for
higher angular momenta requires large  computer
resources. Therefore, in this first study of the 3N continuum with full
N$^3$LO chiral force, we restrict ourselves to low energies only. In
that region of incoming neutron laboratory (lab.) energies below $\sim 30$~MeV,
the most challenging observables are the  nd elastic scattering 
analyzing power and cross
sections in symmetric space star and neutron-neutron
quasi-free-scattering configurations of the nd breakup reaction. The
discrepancies between data and theory for these observables could not
be removed with standard NN and 3NFs \cite{din1}. 

Our paper is organized as follows. 
In Sec. \ref{nucl_ham} we describe our method to determine the
nuclear Hamiltonian by fixing the two parameters $c_D$ and $c_E$ in the
chiral N$^3$LO 3NF. This is achieved by first requiring 
reproduction of the $^3$H binding energy which leads to pairs of 
allowed ($c_D$, $c_E$)
values. Using the experimental data for an additional 3N observable,
which 
in our case
is taken to be the doublet nd scattering length $^2a_{nd}$, fixes
completely the nuclear Hamiltonian at N$^3$LO.  
 Based on the resulting Hamiltonian, we discuss  in Sec.~\ref{nd_elas}  
some results for low-energy elastic 
 nd scattering observables while  in Sec. \ref{nd_breakup} 
 the results for selected low-energy nd breakup configurations are presented. 
 We summarize and conclude in Sec. \ref{summary}.

\section{Determination of nuclear Hamiltonian at N$^3$LO}
\label{nucl_ham}

Neutron-deuteron scattering with neutrons and proton interacting
through  a NN interaction $v_{NN}$ and a 3NF $V_{123}=V^{(1)}(1+P)$, is
described in terms of a breakup operator $T$ satisfying the
Faddeev-type integral equation~\cite{wit88,glo96,hub97}
\begin{eqnarray}
T\vert \phi \rangle  &=& t P \vert \phi \rangle +
(1+tG_0)V^{(1)}(1+P)\vert \phi \rangle + t P G_0 T \vert \phi \rangle + 
(1+tG_0)V^{(1)}(1+P)G_0T \vert \phi \rangle .
\label{eq1a}
\end{eqnarray}
The two-nucleon $t$-matrix $t$ is the solution of the
Lippmann-Schwinger equation with the interaction
$v_{NN}$.   $V^{(1)}$ is the part of a 3NF which is 
symmetric under the interchange of nucleons $2$ and $3$.
 The permutation operator $P=P_{12}P_{23} +
P_{13}P_{23}$ is given in terms of the transposition operators,
$P_{ij}$, which interchange nucleons $i$ and $j$.  The incoming state $
\vert \phi \rangle = \vert \mathbf{q}_0 \rangle \vert \phi_d \rangle $
describes the free nd motion with relative momentum
$\mathbf{q}_0$ and the deuteron state $\vert \phi_d \rangle$.
Finally, $G_0$ is the resolvent of the three-body center of mass kinetic
energy. 
The amplitude for elastic scattering leading to the corresponding
two-body final state $\vert \phi ' \rangle$ is then given by~\cite{glo96,hub97}
\begin{eqnarray}
\langle \phi' \vert U \vert \phi \rangle &=& \langle \phi' 
\vert PG_0^{-1} \vert 
\phi \rangle + 
\langle \phi' \vert PT \vert \phi \rangle + \langle 
\phi'\vert  V^{(1)}(1+P)\vert \phi \rangle
+ \langle \phi' \vert V^{(1)}(1+P)G_0T\vert  \phi \rangle,
\label{eq3}
\end{eqnarray}
while for the breakup reaction one has
\begin{eqnarray}
\langle  \phi_0'\vert U_0 \vert \phi \rangle &=&\langle 
 \phi_0'\vert  (1 + P)T\vert
 \phi \rangle ,
\label{eq3_br}
\end{eqnarray}
where $\vert \phi_0' \rangle$ is the free three-body breakup channel state.

The nuclear Hamiltonian at N$^3$LO  of the chiral expansion  is fixed
by specifying the   
values of LECs $c_D$ and  $c_E$ which parametrize the strengths of
the leading $1\pi$-contact and the three-nucleon-contact terms.  
To determine them we follow the approach of Ref.~\cite{epel2002}
and use the experimental triton binding energy $E({^3H})$ and the 
 nd doublet scattering length $^2a_{nd}$ as two observables from which
$c_D$ and $c_E$ can be obtained. The procedure can be divided into two
steps. First, the dependence of $E({^3H})$ on $c_E$ for a given
value of $c_D$ is determined. The requirement to reproduce the
experimental value of the triton binding energy yields a set of
 pairs $c_D$ and $c_E$. This set is then used in the
calculations of $^2a_{nd}$, which allows us to find the values of
$c_D$ and $c_E$ describing both observables simultaneously. 
As already emphasized above, using the triton binding energy and the
nd doublet scattering length is probably not the optimal way to
fix the parameters in the 3NF due the strong correlation between these
two observables. We will discuss this issue in the next two sections
and present results obtained by relaxing the condition to reproduce
$^2a_{nd}$.

We compute the $^3$H wave function using the method described in
\cite{Nogga1997}, where the full triton wave function 
$\vert \Psi \rangle = (1+P) \vert \psi \rangle$ is
given in terms of its Faddeev component $\psi$, which fulfills the Faddeev
equation
\begin{eqnarray}
\vert \psi \rangle = G_0tP\vert \psi \rangle 
+ (1+G_0t)G_0V^{(1)}(1+P) \vert \psi \rangle .
\label{eq1}
\end{eqnarray}

The doublet scattering length $^2a_{nd}$ is calculated using 
($c_D$,$c_E$) pairs, which reproduce the correct value of
$E({^3H})$. To this end, we solve the Faddeev equation (\ref{eq1a}) 
 for the auxiliary state $ T \vert \phi \rangle $  
at zero incoming energy \cite{zeroenergy}. 
We refer to  \cite{glo96,hub97,book} for a general overview of
3N scattering and for more details on the practical implementation of
the Faddeev equations.

In this first  study, where the full N$^3$LO 3NF is applied, 
 we restrict  ourselves to  nd reactions at low energies, 
$E_{\rm lab, \, n} < 20$ MeV.  
At such low energies it is sufficient to include 
 NN force components with a total two-nucleon angular momenta $j \le 3$ 
 in 3N partial-wave states with the total 3N system angular momentum
 below $J \le 25/2$. For the 3NF it is sufficient  
 to incorporate its matrix elements with $j \le 3$ and $J \le 5/2$. 

Here and in what follows, we employ the N$^3$LO chiral NN potential of 
Ref.~\cite{epel_nn_n3lo,epel_mod}. From among five versions
corresponding to different sets of 
cut-off parameters used to regularize  the Lippmann-Schwinger equation
and in spectral function regularization, namely $(450,500)$~MeV,
$(450,700)$~MeV, $(550,600)$~MeV, $(600,500)$~MeV, and
$(600,700)$~MeV, we applied for the present study two N$^3$LO chiral
NN potentials with cut-off
sets $(450,500)$~MeV and $(450,700)$~MeV, denoted in the following by 201
and 204, respectively. Only for these two sets of cut-offs were we able to
determine   the LECs $c_D$ and
$c_E$ using our procedure. 

\begin{figure}
\includegraphics[scale=0.7]{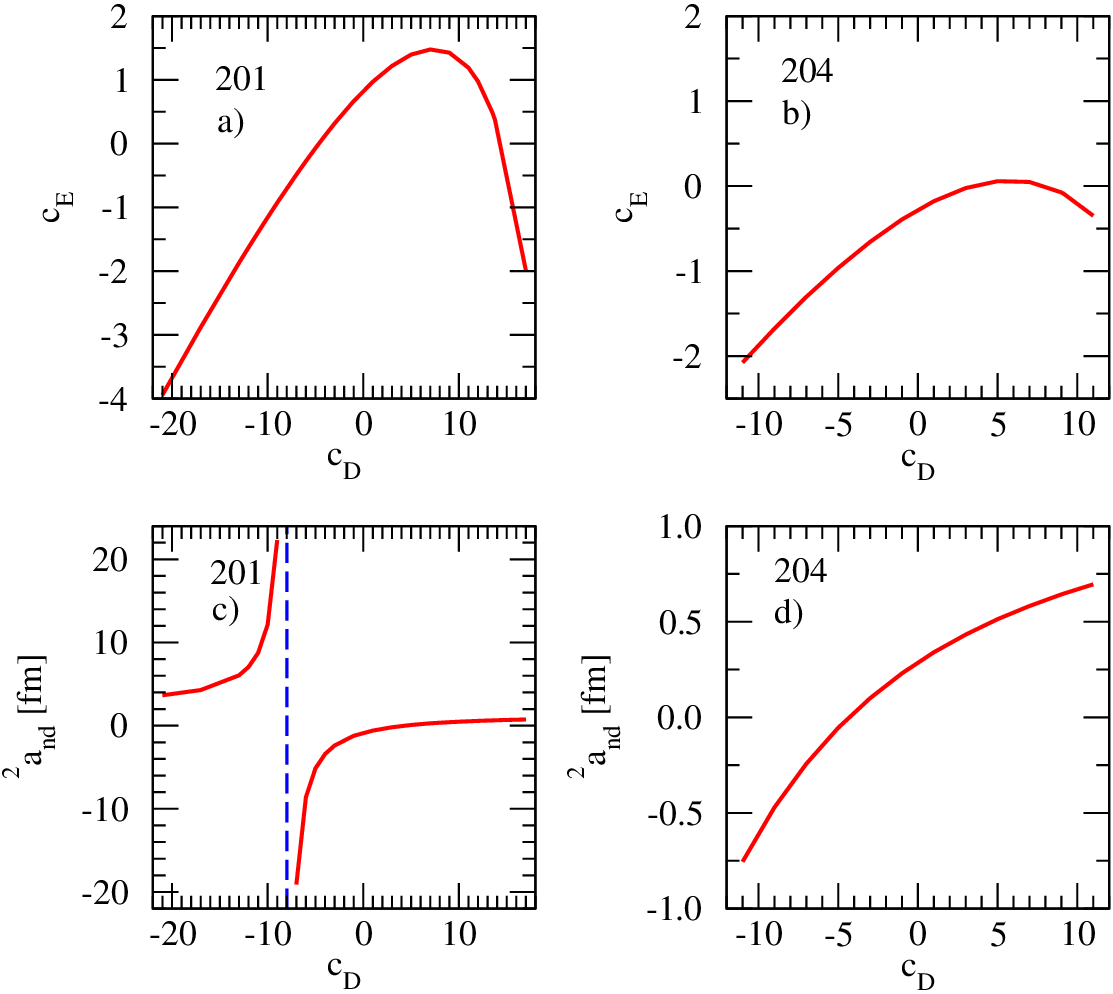}
\caption{
(color online) 
 The dependence of $c_E$ on $c_D$ for versions 201 (a) and 204 (b) of
 the N$^3$LO
 chiral NN Hamiltonian under the condition that the  experimental
 binding energy of $^3$H is reproduced. In c) and d) the corresponding 
 values for doublet nd scattering length are shown. The experimental
 value of the doublet nd scattering length is 
 $^2a_{nd}=0.645(7)$~fm \cite{dandscatt}.
}
\label{fig1}
\end{figure}

In Figs.~\ref{fig1}a and \ref{fig1}b, the sets of $(c_D,c_E)$ values
which reproduce the experimental binding energy of $^3$H are shown, while
in Figs.~\ref{fig1}c and \ref{fig1}d the resulting values of the
doublet nd scattering length $^2a_{nd}$ obtained with such
combinations of $(c_D,c_E)$ are visualized. In the case of the 201 
 N$^3$LO NN chiral   potential a wide range of $c_D$ values have been
 checked and the existence of a pole in the scattering length for
 $c_D \approx -8$ found (see Fig.~\ref{fig1}c). 
 That pole-like behavior reflects the emergence  of an excited state
 for that particular 3N Hamiltonian. 
 The requirement to reproduce, in addition
to the binding energy of $^3$H, also the nd doublet
scattering length leads to the values $(c_D=13.78,c_E=0.372)$ for 201 and
to $(c_D=9.095,c_E=-0.0845)$ for 204 chiral N$^3$LO NN potential. In
the following section we discuss the ambiguities of such a  determination of 
$(c_D,c_E)$. The resulting $c_D$ values are unnaturally large. The
corresponding N$^2$LO values are natural and amount to 
 $(c_D=-0.14,c_E=-0.319)$ and $(c_D=2.43,c_E=0.113)$ for
 $(450,500)$~MeV and  $(450,700)$~MeV
cut-off sets, respectively. 
 It seems that such unnaturally large  values of $c_D$
are not restricted only to the two cut-off sets used in the present study. 
 Namely in \cite{wit_jpg}
 an application of N$^3$LO 3NF, however with  relativistic $1/m$ 
corrections and short-range $2\pi$-contact term omitted, also led to
unnaturally large $c_D$ values for all five cut-off combinations.  
 We hope that new generations of chiral forces with 
 other regularization schemes  will cure this 
 problem \cite{epelbaum_newchiral}.  We also plan to use other 3N
 observables, for example triton $\beta$-decay rate instead of
 $^2a_{nd}$, to fix values of LECs $c_D$ and $c_E$.

\section{Low-energy elastic nd scattering}
\label{nd_elas}

\begin{figure}
\includegraphics[scale=0.9]{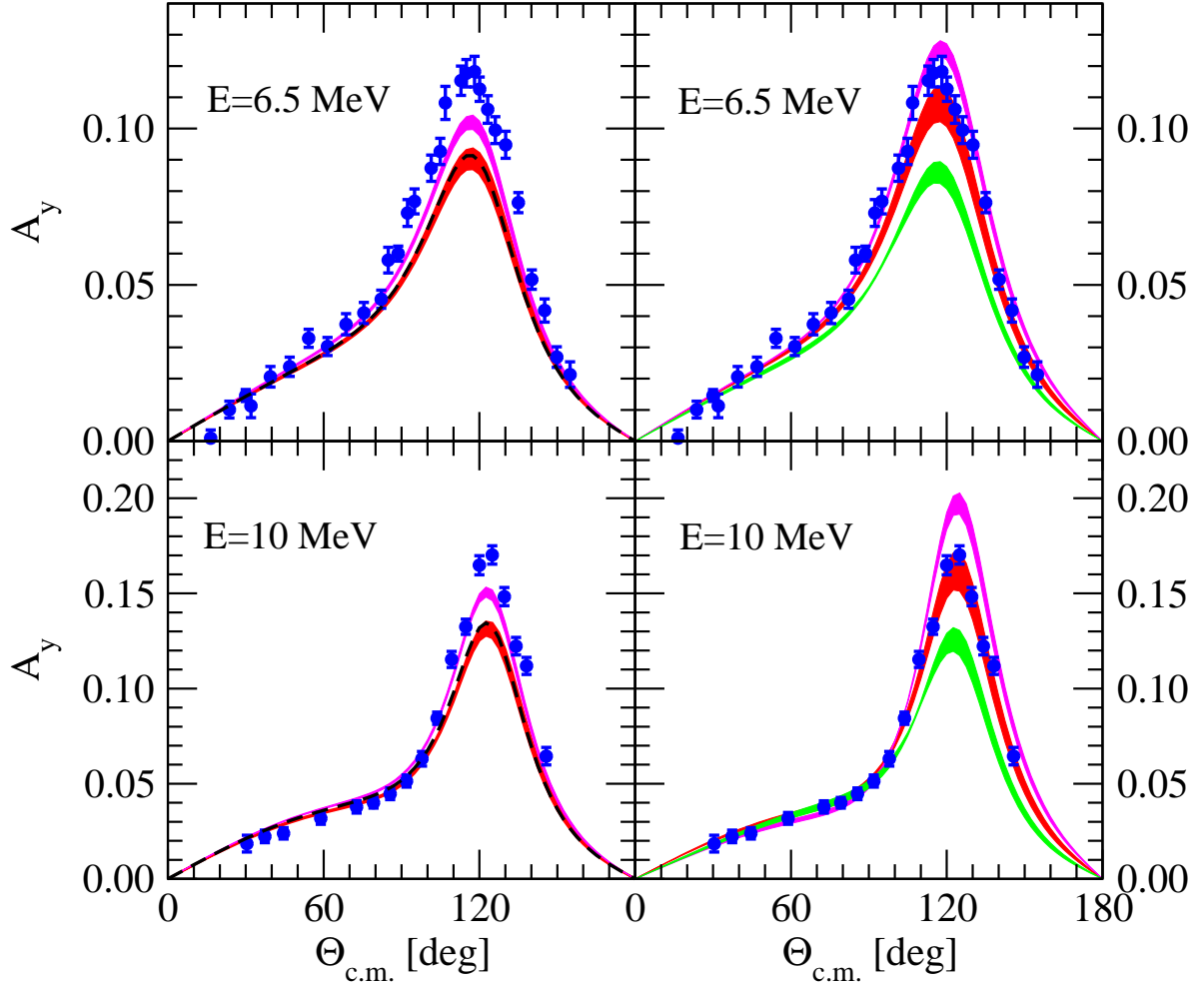}
\caption{
(color online) The nd elastic scattering analyzing power $A_y$  at
  $E_{\rm lab, \, n}=6.5$~MeV and $10$~MeV. 
 In the left panels the bottom  (red) band covers 
  predictions of standard NN potentials: AV18, CD~Bonn, Nijm1 and
  Nijm2. The upper (magenta) band results when these potentials
  are combined with the TM99 3NF. The dashed (black) line shows prediction
  of the AV18+Urbana IX combination. 
 In the right panels bands of predictions for five versions of 
chiral NN potentials at different orders of the chiral expansion are
shown: NLO - the upper (magenta) band, 
 N$^2$LO - the middle (red) band, and N$^3$LO - the bottom 
 (green) band. These five versions correspond to different cut-off
 values used for  the Lippmann-Schwinger equation 
 and spectral function regularizations, namely $(450,500)$~MeV,
$(450,700)$~MeV, $(550,600)$~MeV, $(600,500)$~MeV, and
$(600,700)$~MeV \cite{epel_nn_n3lo,epel_mod}.  
 The full circles are nd data from Ref.~\cite{tornow_ay} at
 $6.5$~MeV and from Ref.~\cite{tornow_ay_10} at $10$~MeV.
}
\label{fig2}
\end{figure}

At low energies of the incoming neutron,
 the most interesting observable is the analyzing
  power $A_y$ for nd elastic scattering with polarized neutrons. 
 Theoretical predictions of standard high-precision NN  
potentials fail to explain the experimental data for $A_y$. The data are 
underestimated 
by $\sim 30 \%$ in the region of the $A_y$ maximum which occurs 
 at c.m. angles $\Theta_{c.m.} \sim 125 ^o$. Combining standard NN 
potentials  
with commonly used models of a 3NF, such as e.g.~the TM99 or Urbana IX models, 
 removes approximately only half of the discrepancy with respect to the 
data (see Fig.~\ref{fig2}). 

When instead of standard forces chiral NN interactions are used, the 
predictions    for $A_y$ vary with the order of chiral expansion
\cite{epel_nn_n3lo,epel_mod}.  In particular, as reported in
Ref.~\cite{epel2002},  the NLO results  overestimate the $A_y$ data 
while N$^2$LO NN forces seem to provide quite a good 
 description of them (see Fig.~\ref{fig2}).  Only when N$^3$LO NN chiral 
forces are used, 
a clear discrepancy between theory and data emerge in the region of $A_y$
 maximum, which is similar to the one for standard forces. This is
 visualized in  Fig.~\ref{fig2}, where bands of predictions for 
five versions of 
the Bochum NLO, N$^2$LO and N$^3$LO potentials with different 
cut-off parameters used for the Lippmann-Schwinger equation 
 and the spectral function
regularizations  are shown \cite{epel_mod}). 
Such behaviour of $A_y$ predictions at different orders in the chiral
expansion  can be traced back to a high sensitivity 
of $A_y$ to $^3P_j$ NN force components and to the fact, that only at N$^3$LO 
of chiral expansion the experimental $^3P_j$ phases 
 \cite{nijm_phase1,nijm_phase2}, 
especially the  $^3P_2$-$^3F_2$ ones, are properly 
 reproduced \cite{wit_jpg}. 
 
\begin{figure}
\includegraphics[scale=0.7]{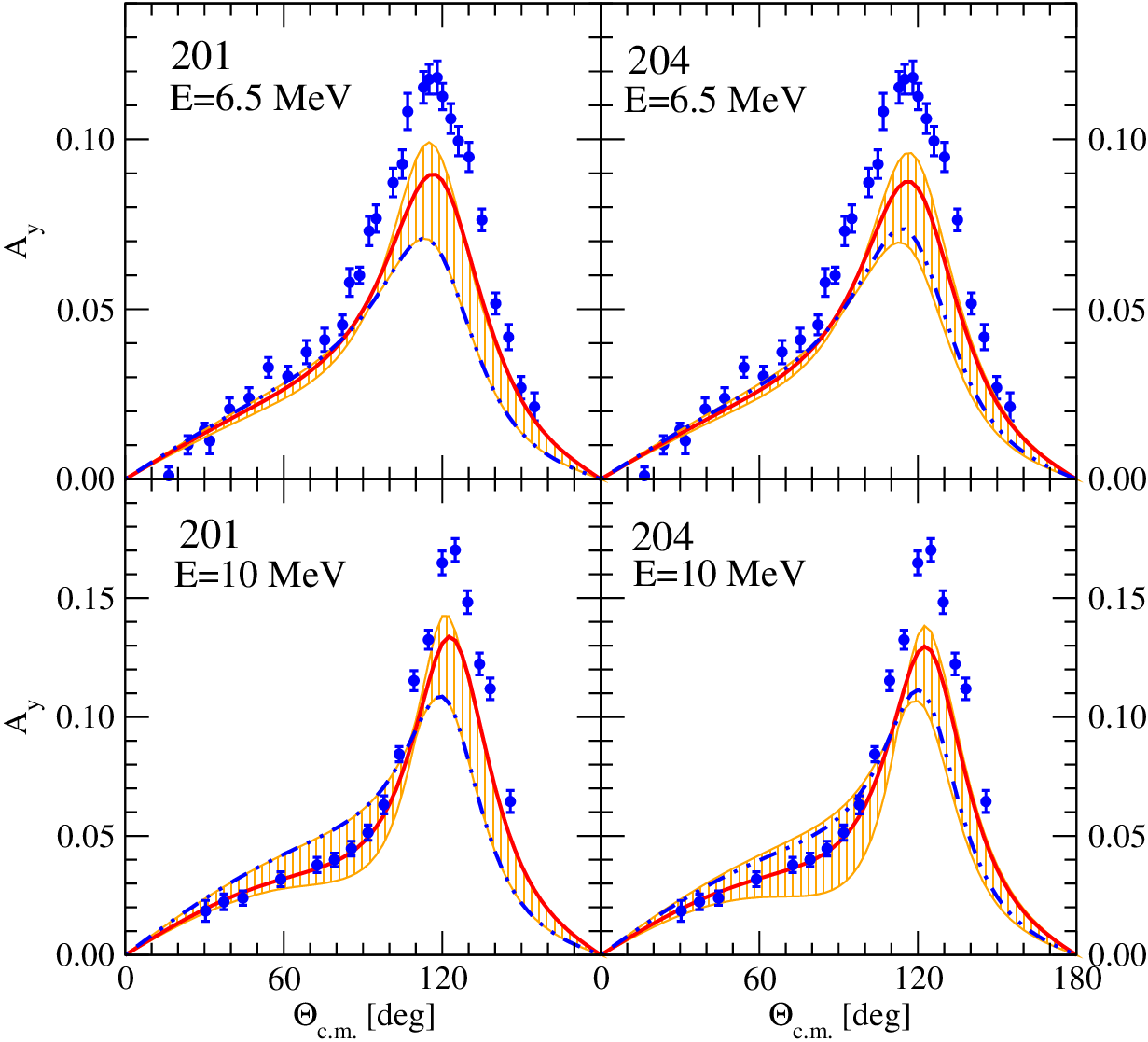}
\caption{
(color online) The nd elastic scattering analyzing power $A_y$  at
  $E_{\rm lab, \, n}=6.5$~MeV and $10$~MeV. The solid (red) lines show
  predictions of the
  N$^3$LO chiral NN potential. The dashed-double-dotted  (blue) 
 lines result when
  the chiral NN potential is combined with the full N$^3$LO 3NF with $c_D$
  and $c_E$ values reproducing  binding energy of $^3$H and
  $^2a_{nd}$ scattering length. The (orange) vertically shaded 
 band covers range of
  predictions for such a combination when pairs of $(c_D,c_E$) values
  from Fig.~\ref{fig1}a and \ref{fig1}b, which reproduce only triton binding
  energy, are used. For the description of the data see Fig.~\ref{fig2}.
}
\label{fig3}
\end{figure}

It is interesting to study whether the consistent chiral N$^3$LO 3NF's 
can explain the low-energy $A_y$-puzzle. In the present
investigation, we, for the first time 
include \emph{all} contributions to N$^3$LO 3NF:  long-range
contributions comprising $2\pi$-exchange, $2\pi-1\pi$-exchange, ring components 
and relativistic  1/m corrections together with short range
 $1\pi$-contact, three-nucleon-contact  and 
  $2\pi$-contact terms. 
 In Fig.~\ref{fig3} we show  by dashed-dotted (blue) line
 the results for   $A_y$ based on the  values of the $c_E$ and $c_D$ 
parameters which reproduce the triton binding energy 
  and $^2a_{nd}$ scattering length.  
It turns out that adding the full N$^3$LO 3NF does not improve 
the description of $A_y$. On the  contrary, adding the chiral N$^3$LO 3NF 
  lowers  the maximum 
of $A_y$ with respect to the chiral N$^3$LO NN prediction, shown by the solid
 (red) line, thus,  
increasing the discrepancy between the theory and the data.

In order to check the restrictiveness of the requirement to reproduce, 
in addition to the $^3$H binding energy, also the 
experimental value of $^2a_{nd}$, we show in Fig.~\ref{fig3} also a band 
of predictions for ($c_E$, $c_D$) pairs from Fig.~\ref{fig1}a and \ref{fig1}b. 
 Even after relaxing the requirement to  reproduce $^2a_{nd}$,  
the $A_y$-puzzle cannot be explained by the N$^3$LO NN and 3NF. 

\begin{figure}
\includegraphics[scale=0.7]{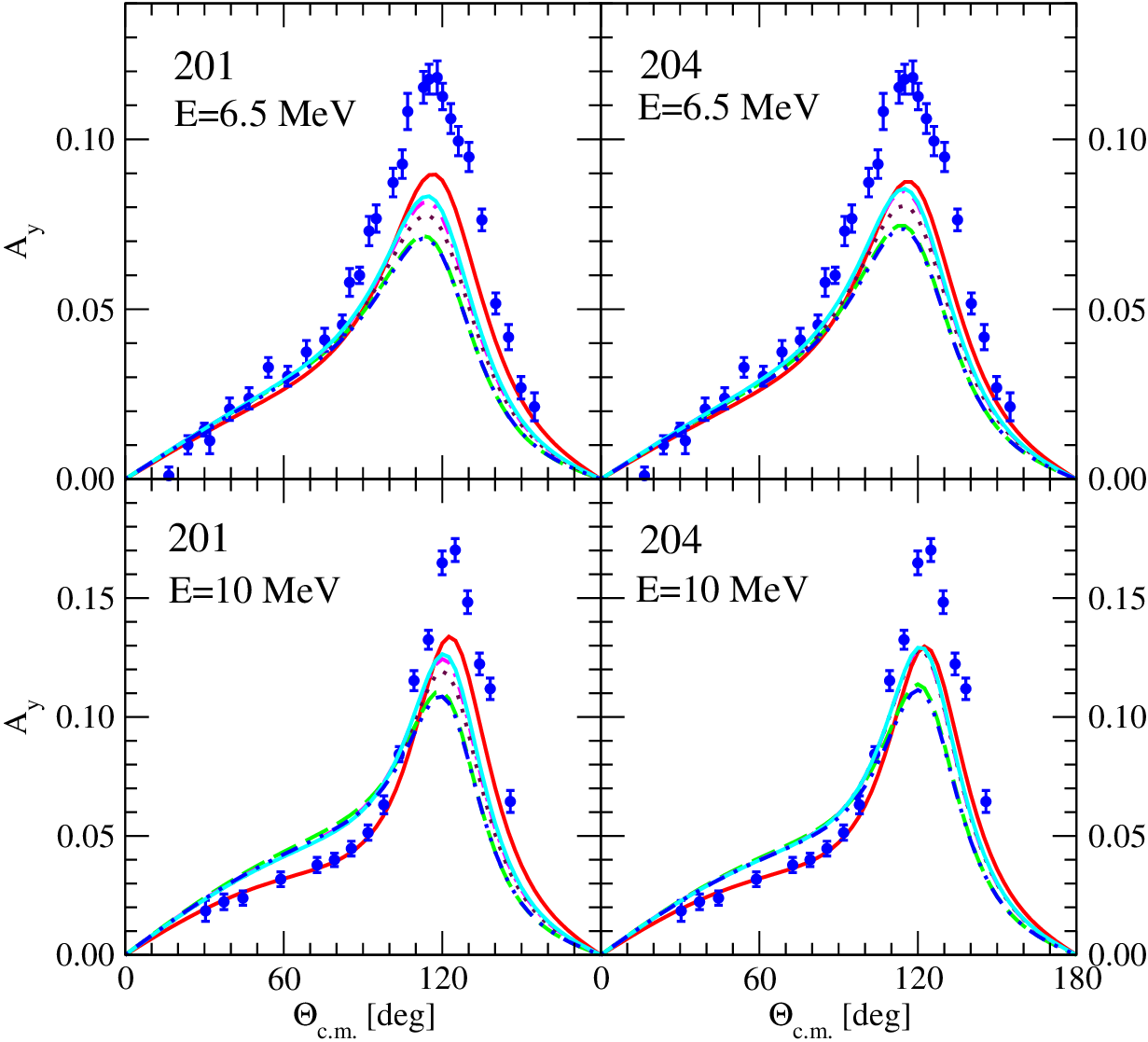}
\caption{
(color online) The nd elastic scattering analyzing power $A_y$  at
  $E_{\rm lab, \, n}=6.5$~MeV and $10$~MeV. The solid (red) line gives the 
prediction of the
  N$^3$LO chiral NN potential. Other lines show the  importance of
  different components of the chiral N$^3$LO 3NF when combined with that
  NN interaction. The solid (cyan), dashed-dotted (magenta), dotted
  (maroon), and dashed (green) lines
  result when that NN N$^3$LO force is combined with $\pi \pi +D+E$,  $\pi \pi
  +D+E+2\pi-contact$, $\pi \pi+ 2\pi1\pi +D+E+2\pi-contact$, 
 and $\pi \pi+ 2\pi1\pi + ring +D+E+2\pi-contact$, respectively. 
 The full N$^3$LO result with the relativistic term included is shown by 
 the dashed-double-dotted (blue) line. 
 For the description of the data see Fig.~\ref{fig2}.
}
\label{fig4}
\end{figure}

It is interesting to see how different components of the N$^3$LO 3NF 
contribute to $A_y$. Taking in addition to the NN N$^3$LO chiral force
only the $2\pi$-exchange term with leading $1\pi$-contact
and three-nucleon-contact terms (these three topologies appear for the
first time
at N$^2$LO) lowers the maximum of $A_y$ (see Fig.~\ref{fig4}, solid
(cyan) line). When, in addition, the short-range $2\pi$-contact component is
included, the value of $A_y$ practically remains unchanged (dashed-dotted
(magenta) line in Fig.~\ref{fig4}). 
This shows that contributions of the $2\pi$-contact term are
 negligible at those energies. 
The long-range $2\pi-1\pi$-exchange and ring
terms lower significantly the maximum of $A_y$ (in Fig.~\ref{fig4} dotted
(maroon) and dashed (green) lines, respectively). Finally, inclusion
of the relativistic $1/m$ contribution leaves the maximum of $A_y$
 practically unchanged (dashed-double-dotted (blue) line in Fig.~\ref{fig4}). 
 It  should be pointed out that when taking into account the $1/m$
 corrections to the N$^3$LO 3NF, one should also include the
 corresponding relativistic corrections in the NN force  and, in
 addition, also relativistic corrections to the kinetic energy, which
are  formally of the same importance. This would considerably
complicate the calculation.  In our present work, we do not take into
account such corrections and employ the standard nonrelativistic
framework. This seems to be justified in view
 of the low energies considered in this paper and the very small effects
 caused by relativistic $1/m$ corrections to the 3NFs found in this
 study. Last but not least, we emphasize that the contributions of the
 individual 3NF topologies to the  $A_y$ puzzle are not observable and
 depend, in particular, on the regularization scheme and employed NN
 forces.

\begin{figure}
\includegraphics[scale=0.7]{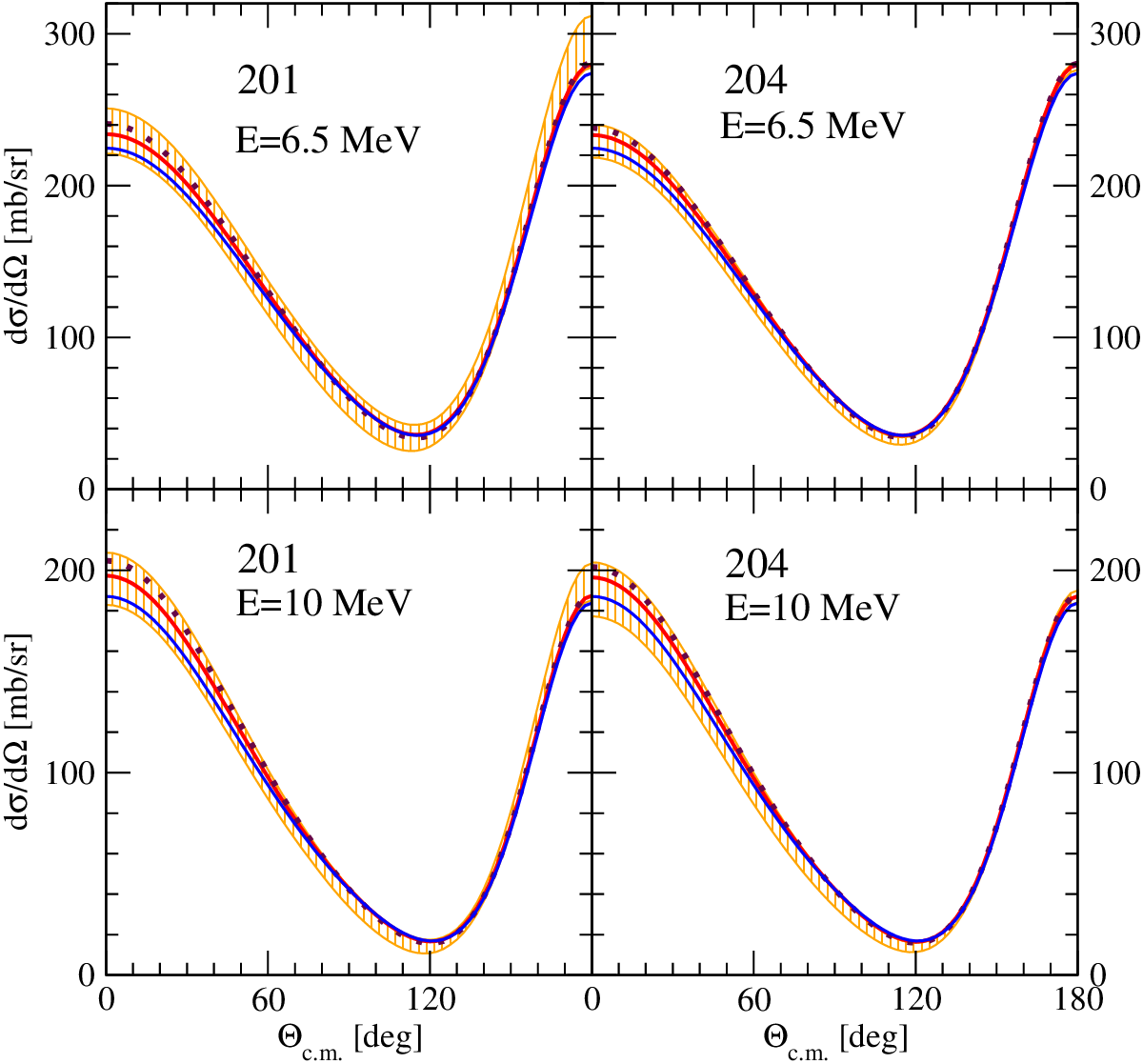}
\caption{
(color online) The nd elastic scattering angular distributions at
  $E_{\rm lab, \, n}=6.5$~MeV and $10$~MeV. 
 The solid (blue) lines show predictions of the
  CD~Bonn potential. The solid (red) lines give predictions of the
  N$^3$LO chiral NN potential. The dotted (maroon) lines result when
  the chiral N$^3$LO NN potential is combined with full N$^3$LO 3NF with $c_D$
  and $c_E$ values reproducing both binding energy of $^3$H and
  $^2a_{nd}$ scattering length. The (orange) vertically shaded 
 band covers the range of
  predictions for such a combination when pairs of $(c_D,c_E$) values
  from Fig.~\ref{fig1}a and \ref{fig1}b, which reproduce only triton binding
  energy, are used.
}
\label{fig5}
\end{figure}

It is important to address the question of uniqueness of our approach 
to determine the constants $c_D$ and $c_E$. To this aim, we checked 
how taking  instead  of $^2a_{nd}$ 
 a different nd observable 
 would influence determination of 
 $c_D$ and $c_E$. The low-energy elastic nd scattering cross section
is an observable which seems to be reasonably well described by
standard theory \cite{calvin_nd}. In Fig.~\ref{fig5} we show (orange) bands of
predictions for the nd elastic scattering cross section at
$E_{\rm lab, \, n}=6.5$~MeV and $10$~MeV obtained with full N$^3$LO chiral force 
 with ($c_D$, $c_E$) values from Figs.~\ref{fig1}a and \ref{fig1}b which
 reproduce only the experimental binding energy of $^3$H. These  
bands are relatively narrow for version 204 and 
 angles ${\Theta}_{c.m.}> 130^{\circ}$
and start  to become broader at smaller angles. At forward angles the
requirement that only the binding energy of $^3$H is reproduced leads to
 a wide range of predictions for the cross section. The solid (red) lines
in Fig.~\ref{fig5} are predictions of the N$^3$LO chiral NN potential and the
  dotted (maroon) lines show cross sections for the full N$^3$LO chiral
  force with constants  $c_D$ and $c_E$ fixed  by requirement
  that the doublet nd $^2a_{nd}$ scattering length is also reproduced. 
 For comparison to standard potential cross sections  in
  Fig.~\ref{fig5} also the CD~Bonn potential results are shown by
  solid (blue) lines. 
 The backward angle nd elastic scattering cross section data  are
 properly described by standard, high precision NN potentials \cite{calvin_nd}. 
  To fix values of $c_D$ and $c_E$ it would be desirable to have
  forward angle cross section data.  Assuming that in
  this angular region the data will be properly described by our
  theory indicates that replacing $^2a_{nd}$ by cross section 
 would lead to
  consistent   $c_D$ and $c_E$ values in both approaches.

\section{Low-energy nd breakup}
\label{nd_breakup}

Among numerous kinematically complete configurations of the nd breakup 
reaction the SST and QFS configurations have attracted special
attention. 
 The cross sections for these geometries 
are very stable with respect to the underlying dynamics.
Different potentials, alone or combined
with standard 3NFs, lead to very similar results for the 
cross sections \cite{din1} which deviate significantly from available
 SST and neutron-neutron (nn) QFS data.  
 At low energies,  the cross sections in the SST and QFS configurations are
 dominated by the S-waves. For the SST configuration, the largest
contribution to the cross section comes from the $^3S_1$ partial 
 wave, while for the nn QFS
 the $^1S_0$ partial wave dominates.
Neglecting rescattering, the QFS configuration resembles free NN
scattering. For free, low-energy neutron-proton (np) scattering one expects
contributions from $^1S_0$ np and $^3S_1$ force components. For free nn
scattering, only the $^1S_0$ nn channel is allowed. This suggests that
the nn QFS
is a powerful tool to study the nn interaction.
The measurement of np QFS cross sections  have revealed good agreement
 between the data and theory \cite{exqfs}, thus confirming the knowledge of
 the np force.
For the nn QFS it was found that the theory underestimates the data by
$\sim 20\%$ \cite{exqfs}. The large
stability of the QFS cross sections
 with respect to the underlying dynamics means that, assuming
 correctness of the nn QFS data,  the present day
 $^1S_0$ nn interaction is probably incorrect \cite{din1,din2,din3}. 

\begin{figure}
\includegraphics[scale=0.8]{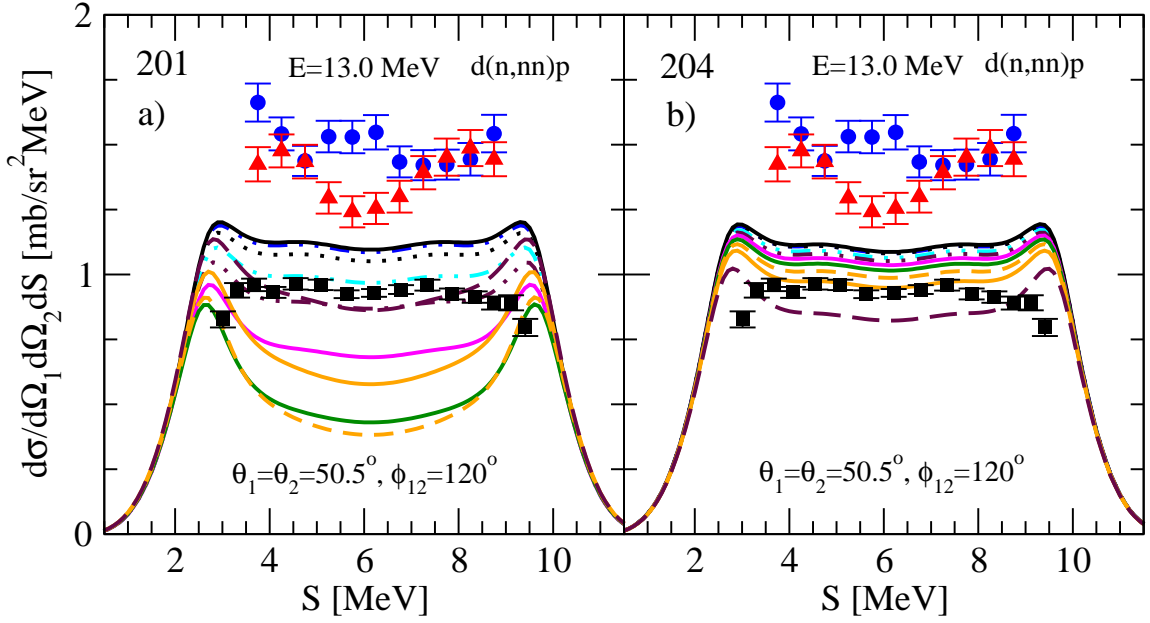}
\caption{ 
(color online) The SST nd breakup cross
  section at incoming neutron lab. energy $E_{\rm lab, \, n}=13$~MeV as a
  function of the arc-length S along the kinematical locus in the
  $E_1-E_2$ plane.  
  The solid (black) line shows the prediction of the chiral N$^3$LO NN potential
  alone. 
 The dashed-double-dotted (blue) line results when the full N$^3$LO chiral 3NF
 with $c_D$ and $c_E$ values reproducing binding energy of $^3$H and
 $^2a_{nd}$ is added to the chiral N$^3$LO NN potential. 
 Other lines show the result when that NN force is
  combined  with the full N$^3$LO 3NF with pairs ($c_D$,$c_E$)
  reproducing only  the experimental binding energy of $^3$H 
  from Fig.~\ref{fig1}a for a) [Fig.~\ref{fig1}b for b)]: 
(9.0,1.425) [(9.0,-0.0752)] - (black) dotted, 
(5.0,1.395) [(5.0,0.058)] - (cyan) dashed-double-dotted,
(3.0,1.219) [(3.0,-0.023)] - (maroon) dashed-double-dotted), 
(1.0,0.971) [(1.0,-0.178)] - (magenta) solid, 
(-1.0,0.6655) [(-1.0,-0.392)] - (green) solid, 
(-3.0,0.3155) [(-3.0,-0.656)] - (orange) dashed, 
(-5.0,-0.071) [(-5.0,-0.962)] - (orange) solid, 
(-9.0,-0.92883) [(-9.0,-1.6759)] - (maroon) dashed. 
  The (blue) circles and (red) triangles are  nd data from
  Ref.~\cite{sst} and \cite{erlangen13a,erlangen13b}, respectively. 
The (black) squares are proton-deuteron (pd) data of Ref.~\cite{koln13}.
}
\label{fig6}
\end{figure}

Also the  chiral
 N$^3$LO forces with all components of the 3NF included 
 are not an exception and  cannot explain the discrepancy 
 between the theory 
and data found for the SST configuration \cite{sst} (Fig.~\ref{fig6}).
  The solid (black) line shows the cross section when only NN
chiral N$^3$LO force is active. Adding the full N$^3$LO 3NF with $c_D$ and
$c_E$ pairs reproducing the experimental binding energy of
$^3$H and nd doublet scattering length leads to
dashed-double-dotted (blue) line. At $13$~MeV, it lies only slightly
below the NN potential prediction indicating only small 3NF effects at
this energy.

It is interesting to see how the SST  cross section depends on the 
choice of parameters ($c_D$,$c_E$) which enter the N$^3$LO nuclear
Hamiltonian. 
  In Fig.~\ref{fig6}, the  SST
cross sections at $E_{\rm lab, \, n}=13$~MeV  are shown for a number
of $c_D$ and $c_E$ pairs which reproduce only the experimental binding
energy of $^3$H (taken from Fig.~\ref{fig1}a and \ref{fig1}b). 
 For the 201 N$^3$LO nuclear Hamiltonian (see Fig.~\ref{fig6}a)  
 decreasing the value of $c_D$ 
 leads to big changes of
 the SST cross section. 
 Starting from $c_D=13.78$, which reproduce also $^2a_{nd}$, and
 decreasing it to $c_D=9$ leads to only small changes of the SST cross
 sections. Further lowering of $c_D$ down to $c_D=-3$ reduces  the
 cross section 
 and the discrepancy to nd data at $13$~MeV is drastically
increased. If we continue to reduce the  $c_D$ value  the  SST cross
section rises, however, it remains always below the pure NN prediction. For the
204 N$^3$LO nuclear Hamiltonian the changes of the SST cross section
are not so drastic and decrease of the $c_D$ reduces the cross
section (see Fig.~\ref{fig6}b).   
 Thus, in spite of the strong sensitivity of the SST cross sections to
 values of $c_D$ and $c_E$, it is not possible to describe the
 available experimental
data for the  SST nd cross sections at $13$~MeV even allowing for
pairs of ($c_D$,$c_E$) which do not reproduce $^2a_{nd}$. 

\begin{figure}
\includegraphics[scale=0.8]{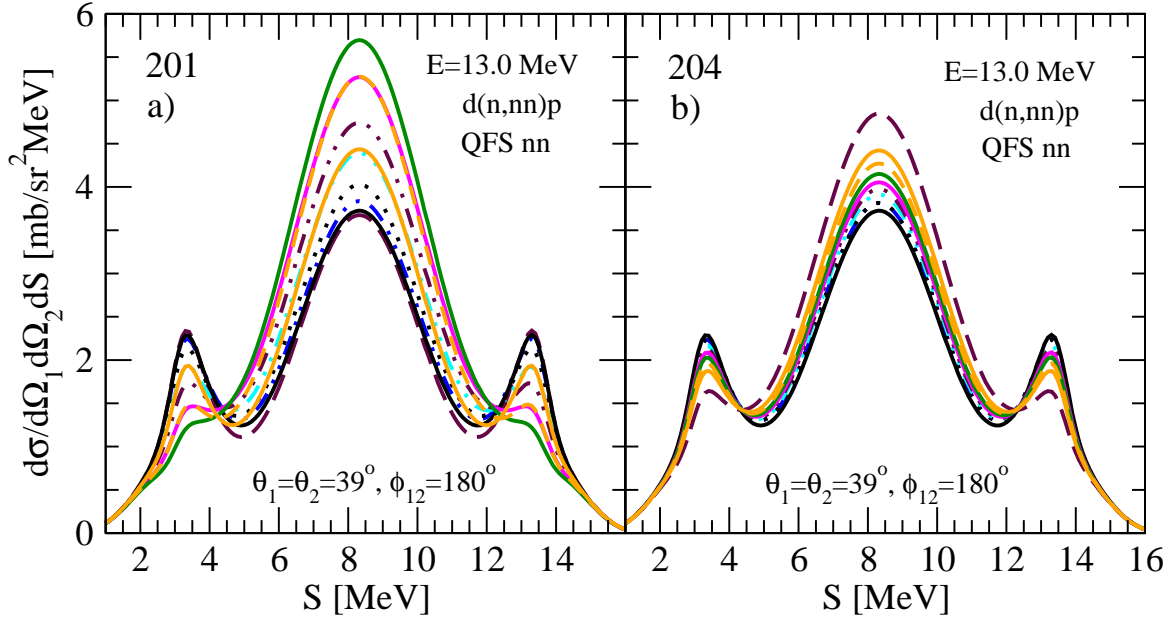}
\caption{
(color online)  The nn QFS nd breakup cross
  section at incoming neutron lab. energy $E_{\rm lab, \, n}=13.0$~MeV.
 For an explanation of the lines see Fig.~\ref{fig6}.
}
\label{fig7}
\end{figure}

As shown in Fig.~\ref{fig7} the behaviour of the QFS cross section 
is different from SST. This configuration  also appears to be
sensitive to changes of
$c_D$ and  $c_E$ values. Here, decreasing $c_D$  for the 201 N$^3$LO
nuclear Hamiltonian leads first to the increase
of the QFS cross section up to $c_D \sim -1.0$. Further lowering 
the value  
of $c_D$ reduces the QFS cross section (see Fig.~\ref{fig7}a). 
 For the 204 N$^3$LO
nuclear Hamiltonian decreasing $c_D$  leads  to the increase
of the QFS cross section (see Fig.~\ref{fig7}b). 
 The values of $c_D$ and  $c_E$ 
 which reproduce the $^3$H binding energy and $^2a_{nd}$
 lead only to a slight increase of the QFS cross 
section with respect to the N$^3$LO
NN prediction and thus to small 3NF effects.

\section{Summary and outlook}
\label{summary}

Recent efforts towards the derivation and implementation of the
N$^3$LO 3NF allowed us, for the first time, to apply the full chiral 
 N$^3$LO Hamiltonian 
to the low-energy nd elastic scattering 
 and breakup reactions. The nuclear Hamiltonian at that order of the chiral
 expansion is unambiguously given after fixing the two constants $c_D$ and $c_E$
 which determine the strengths of the $1\pi$-contact and
 three-nucleon-contact components of the N$^3$LO chiral 3NF. 
 We determined these low-energy constants by requiring reproduction of 
 the  binding energy of
 $^3$H and the doublet nd scattering length $^2a_{nd}$. 
 We found indications that using low-energy
 nd  elastic scattering cross section instead of $^2a_{nd}$ would
 probably lead to similar values of these parameters. 
 
 It turns out that applying the full N$^3$LO 3NF with specific cut-off
 parameters used in this study cannot explain 
 the low-energy $A_y$-puzzle. Contrary to the 3NF effects found for $A_y$
 with standard NN potentials combined with 3NF models such as TM99 or
 Urbana IX, where the inclusion of the  3NF decreased the
 discrepancy to data by about $\sim 50 \%$, the chiral N$^3$LO 3NF
combined with the NN potential of 
 Ref.~\cite{epel_nn_n3lo} lowers the maximum of $A_y$ 
 increasing the discrepancy. 
 It should, however, be emphasized that the low-energy 3N $A_y$ is a fine-tuned
 observable which is very sensitive to changes in $^3P_j$ NN force
 components as well as to P-waves in the Nd system
 \cite{hub1995,tor1998}. Thus, the disagreement with the data must be 
  interpreted with considerable caution.  
 Our result suggests the lack of some spin-isospin-momenta  structures
 in the N$^3$LO 3NF. However, possible inaccuracies in low-energy $^3P_j$ NN 
phase-shifts cannot be excluded. 
 The 3NF derived in the standard formulation of 
chiral perturbation theory 
based on pions and nucleons as the only explicit degrees of freedom is
known to  miss certain significant intermediate-range
contributions of the $\Delta$(1232)  resonance at N$^3$LO, 
 which, to some extent, are accounted for only at N$^4$LO and higher 
orders \cite{krebs1,krebs2}. 
 It would therefore be interesting, to apply the recently 
 derived N$^4$LO 3NF \cite{krebs1,krebs2} in calculations of nd
 reactions together with subleading contributions to the
   three-nucleon contact interactions \cite{girl_2011}. The
   short-range 3N forces at N$^4$LO which contribute to Nd P-waves
   may solve the $A_y$-puzzle in a trivial way. 

We found that cross sections in kinematically complete SST and QFS nd
breakup configurations at low energies are quite sensitive to the 
values of $c_D$ and $c_E$. For
 their  values fixed by the experimental 
  binding energy of $^3$H and $^2a_{nd}$ 
only small 3NF effects were found in these
configurations. Large discrepancies with the data remain in these
configurations.

For the SST geometry at $13$~MeV, there is a serious discrepancy
between theory and two independent nd data sets of
Refs.~\cite{sst,erlangen13b} as well as between theory and
proton-deuteron (pd) data of
Ref.~\cite{koln13}. While the nd data lie  $\sim 20 \%$ above the
theory, the pd data lie $\sim 10 \%$ below theory and 
$\sim 30 \%$  below nd data. Recent pd calculations with Coulomb
force included show practically negligible effects of the
proton-proton Coulomb force for this configuration \cite{deltuva}. The
observed large splitting between the nd and pd data indicates either
that there are large isospin-breaking effects or that the data are not
consistent. 

Higher-energy nd reactions, in which clear evidence of large 3NF effects 
 was found, call for applications of the full N$^3$LO force. 
 Studies of the cut-off dependence of  
N$^3$LO NN chiral interaction in higher-energy nd elastic scattering 
 revealed preference for larger cut-off values \cite{wit_jpg}. The use of 
 lower cut-offs would 
preclude applications of N$^3$LO chiral dynamics in that interesting region 
of energies. 
It is important to address the issue of reducing finite-cutoff
artifacts and increasing the
accuracy of chiral nuclear forces prior to 
applying the  chiral N$^3$LO Hamiltonian
at higher energies. 
 In addition, one needs to explore different possibilities to
determine the
LECs entering the 3NF in view of the known strong correlations
between e.g.
the $^3$H and $^4$He binding energies and the nd doublet scattering
lengths, see
 \cite{Gazit:2008ma} for a related discussion. Last but not least, more effort
should be invested  into providing a reliable estimation of the
theoretical uncertainty
at a given order in the chiral expansion. 
 Work along these lines is in progress.

\section*{Acknowledgments}
This study has been performed within Low Energy Nuclear Physics
International Collaboration (LENPIC) project and 
was  supported by the Polish National Science Center 
 under Grant No.DEC-2013/10/M/ST2/00420. 
 It was also supported in part  
 by the  European Community-Research Infrastructure
Integrating Activity
``Exciting Physics Of Strong Interactions'' (acronym WP4 EPOS)
under the 
Seventh Framework Programme of EU, the ERC project 259218
NUCLEAREFT, by the Foundation for Polish Science MPD program,
cofinanced by the European Union within the
Regional Development Fund, by the US Department of Energy under
 Grant Nos. DESC0008485 (SciDAC/NUCLEI) and DE-FG02-87ER40371, by the
 US National Science Foundation under Grant No. PHYS-0904782, and by
 the ERC Grant No. 307986 STRONGINT.  
 The numerical calculations have been performed on the supercomputer 
clusters of the JSC, J¨ulich, Germany, the Ohio Supercomputer Centre, 
USA (Project PAS0680) and the Argonne Leadership Computing Facility
(ALCF) at Argonne National Laboratory 
 (Resource Project:   NucStructReact), where an award of computer time 
was provided by the Innovative and Novel Computational Impact on 
Theory and Experiment (INCITE) program. This research used 
resources of the ALCF, which is supported by the Office of Science of 
the U.S. Department of Energy under contract DE-AC02-06CH11357.

\newpage 
\newpage
\newpage
\newpage
\newpage
\newpage
\newpage
\newpage

\end{document}